\documentclass[aps,asmsymb,twocolumn,showpacs,amsmath,prl]{revtex4-1}

\usepackage{graphicx}
\usepackage{subfigure}
\usepackage{epsfig}
\usepackage{dcolumn}
\usepackage{bm}
\usepackage{color}
\usepackage{xcolor}
\usepackage{bm}
\usepackage{hyperref}
\usepackage{amsmath, amssymb, dsfont}

\def\be{\begin{equation}}       \def\ee{\end{equation}}
\def\bea{\begin{eqnarray}}      \def\eea{\end{eqnarray}}
\def\ba{\begin{array}}
\def\ea{\end{array}}
\def\bnum{\begin{enumerate} }
\def\enum{\end{enumerate}}

\def\=>{\Rightarrow}
\def\>{\rightarrow}

\def\eye2{Fathbb{I}}
\def\bk{{\bf k}}

\def\bK{{\bf K}}
\def\bS{{\bf S}}

\def\si{{\sigma}}

\def\bS{{\bf S}}
\def\bL{{\bf L}}

\renewcommand{\>}{\rangle}

\newcommand{\eps}{\epsilon}

\newcommand{\al}{\alpha}

\begin{document}

\title{Coexistence of spin-1 fermion and Dirac fermion on the triangular kagome lattice}
\author{Luyang Wang}
\author {Dao-Xin Yao}
\email{yaodaox@mail.sysu.edu.cn}
\affiliation{State Key Laboratory of Optoelectronic Materials and Technologies, School of Physics, Sun Yat-Sen University, Guangzhou 510275, China}

\begin{abstract}
Quasiparticle excitations beyond Dirac-Weyl-Majorana classification can appear in lattice systems due to the less symmetry constraint compared with Poincar\'e symmetry in high energy physics. In particular, fermions with an integer spin can appear in a variety of lattices. Here, we show that two-dimensional spin-1 fermion may coexist with Dirac fermions in the triangular kagome lattice (TKL). We derive a four-band effective model that hosts both types of fermions. The effective model can be used to study the interplay between spin-1 and spin-1/2 fermions. As an example, using this model we show that spin-nonconserving Klein tunneling can occur in the TKL, which has the transmission coefficient $T=1$ for normal incidence. Our findings pave a way to the study of the interaction and interplay between different types of fermions in lattice systems.
\end{abstract}
\date{\today}
\maketitle	

{\it Introduction}.---Fermionic quasiparticles which have no counterpart in particle physics can emerge in condensed matter systems \cite{Bradlyn2016}. In particle physics, Poincar\'e symmetry constrains the types of fermions, and they are classified by Dirac-Weyl-Majorana regime\cite{Pal2011}. While Dirac fermions are found, the existence of Weyl fermion and Majorana fermion is still under debate. In condensed matter systems, lattices respect space group symmetries and are less constrained, hence they can host more types of fermions. Besides Dirac and Weyl fermions, fermions with higher spin including spin-1 and spin-3/2\cite{Bradlyn2016} and other types such as triple point fermion\cite{Zhu2016,Weng2016a} could appear in three dimensional solids, and are protected by space group symmetries.

Fermions with higher spin could also appear in two dimensional (2D) lattices, which has been shown in a variety of cases\cite{Bercioux2009,Apaja2010,Green2010,Shen2010,Urban2011,Dora2011,Lan2011,Vigh2013, Xu2014,Romhanyi2015,Zhu2017,Essafi2017}. In particular, spin-1 fermions may appear in $\mathcal{T}_3$ lattice\cite{Bercioux2009}, Lieb lattice\cite{Apaja2010,Shen2010}, kagome lattice\cite{Green2010,Essafi2017} and stacked triangular lattice\cite{Dora2011}. Spin-1 fermions in 2D exhibit distinct novel effects, such as super-Klein tunneling\cite{Urban2011,Xu2014}, diverging dc conductivity\cite{Vigh2013} and unconventional quantum Hall effect\cite{Xu2017}.

To study the interplay between different types of fermions which can be quite interesting, systems with their coexistence is highly desired. Such coexistence has been found in several 3D systems. For example, it has been shown that Weyl fermions coexist with triple point fermions in ZrTe\cite{Weng2016b}; Weyl fermions and spin-3/2 fermions coexist in tricolor cubit lattice\cite{Ezawa2016}; and Weyl fermions coexist with spin-1 or spin-3/2 fermions in transition metal silicides\cite{Tang2017}.

In this work, we investigate the fermionic quasiparticle excitations in the 2D triangular kagome lattice (TKL). While the magnetism on the TKL has been studied both experimentally\cite{Gonzalez1993,Maruti1994,Mekata1998} and theoretically\cite{Loh2008a,Yao2008,Loh2008b,Chen2012,Kim2017}, the band properties of noninteracting fermions on the TKL have not been well studied. Here, we demonstrate that in the TKL, spin-1 fermion and Dirac fermion can coexist. Using $k\cdot p$ theory, we derive a four-band effective model that hosts both types of fermion. The effective model provides a good start to study the interplay between the two types of fermions. As an example, we study the Klein tunneling of a fermion from the spin-1 particle state to the spin-1/2 hole state using the effective model. We find that in this process of spin-nonconserving Klein tunneling, the transmission coefficient is $T=1$ for normal incidence, the same as the Klein tunneling of Dirac fermion and spin-1 fermion.

\begin{figure}[t]
  \centering
  \subfigure[]{\includegraphics[width=5cm]{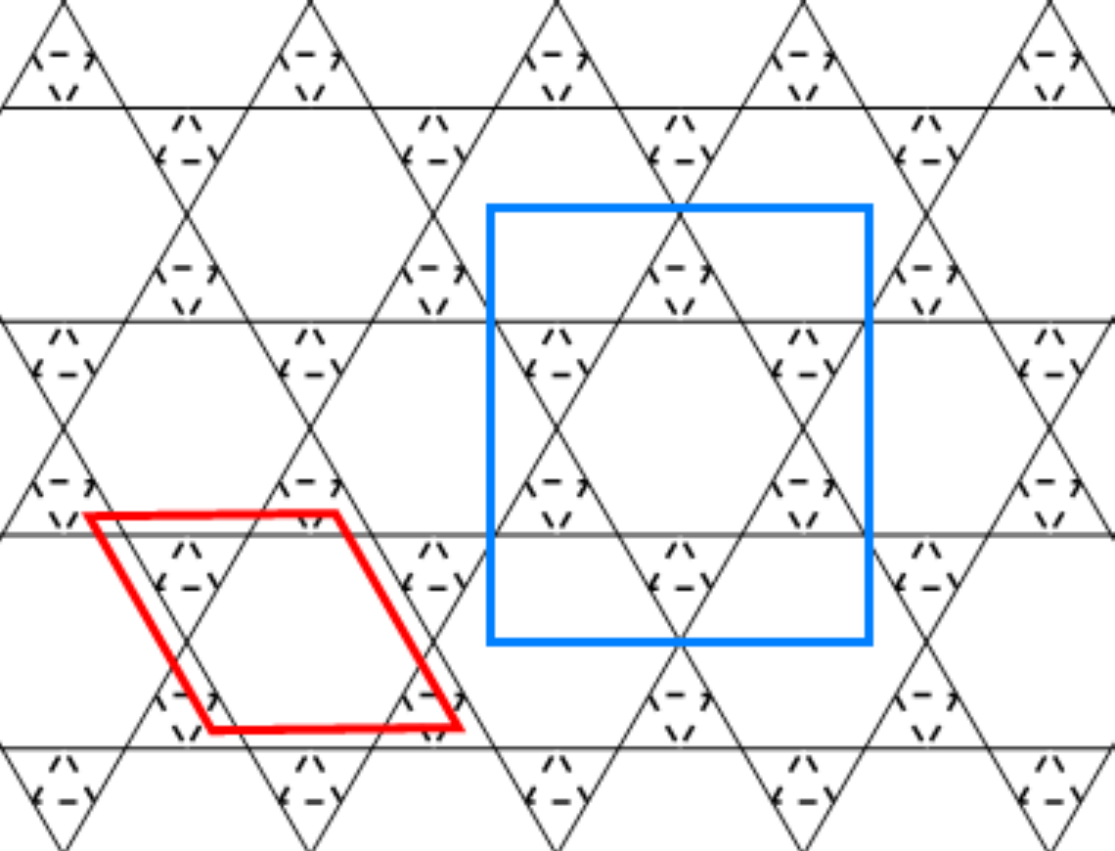}\label{Lattice}}~~
  \subfigure[]{\includegraphics[width=3.cm]{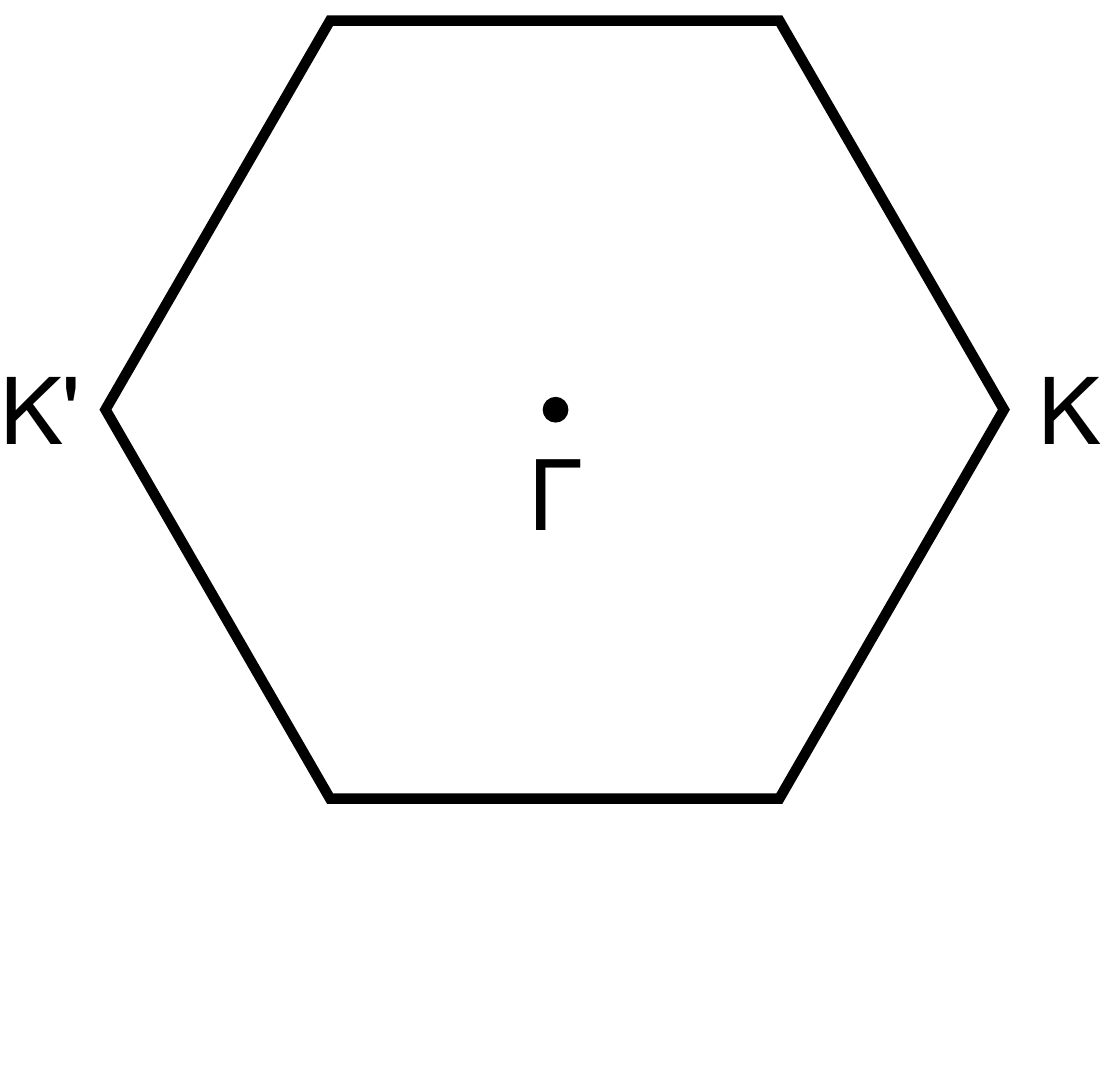}\label{BZ}}
  \caption{(a) The triangular kagome lattice (TKL). Solid lines and dashed lines represent the hopping amplitude $t$ and $t'$, respectively. A unit cell contains nine sites, as indicated by the red parallelogram. From the part enclosed by the blue rectangle, one can see the TKL has $D_{6h}$ point group symmetry. (b) The first Brillouin zone with the high symmetry points labeled.}\label{Fig1}
\end{figure}

\begin{figure}[t]
  \centering
  \subfigure[]{\includegraphics[width=4cm]{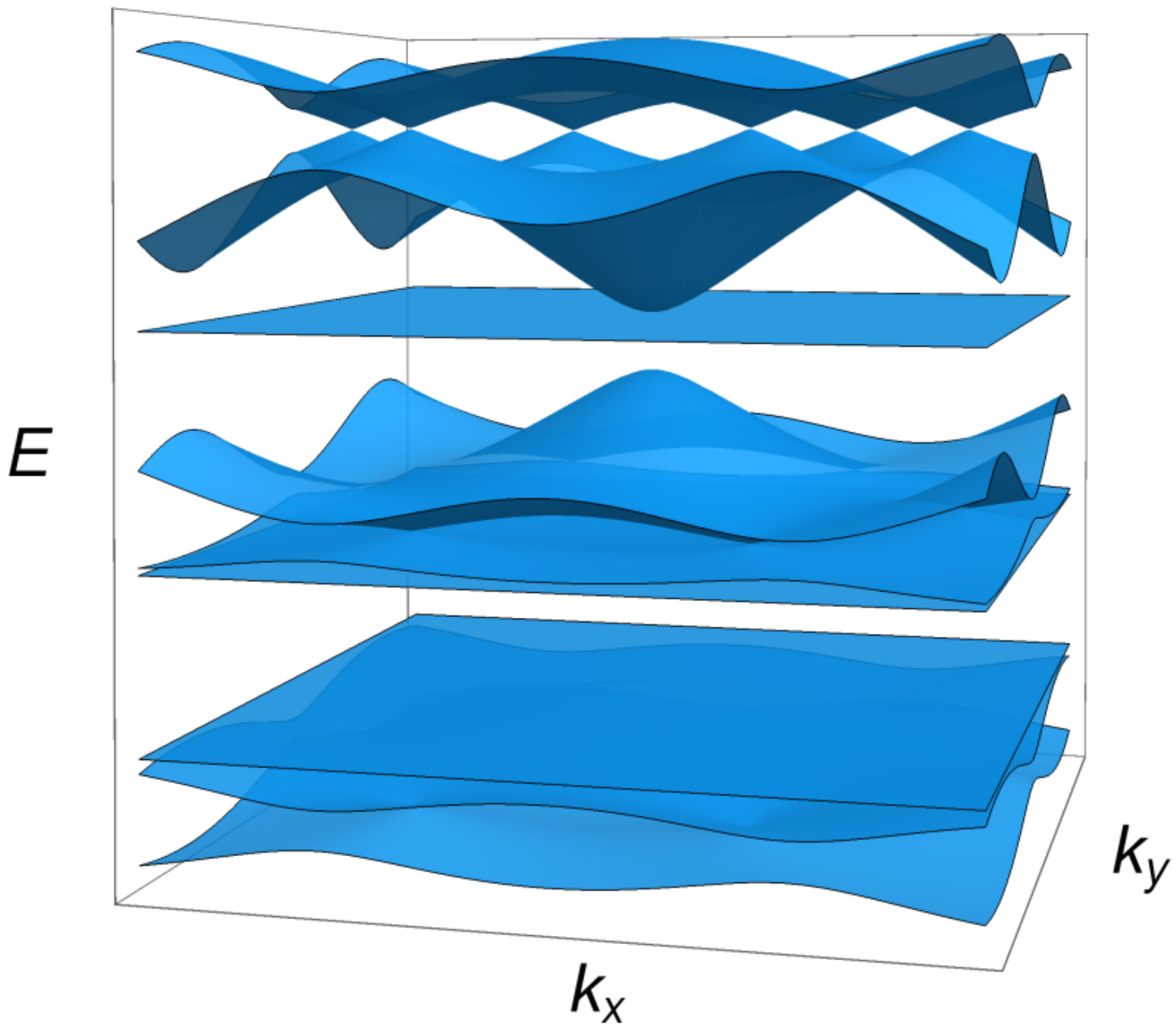}\label{Ek}}~~
  \subfigure[]{\includegraphics[width=4.cm]{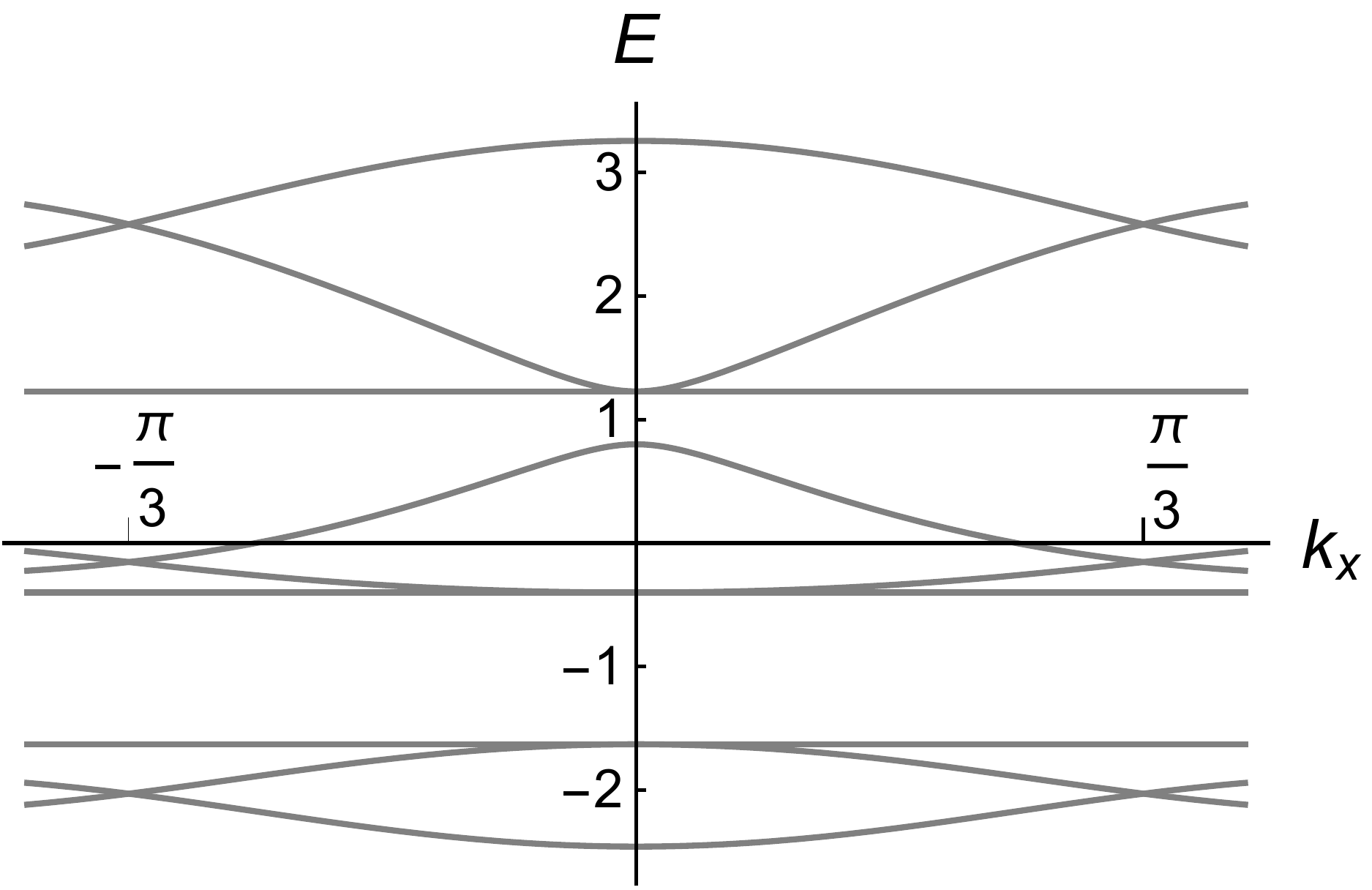}\label{Ekx}}
  \caption{(a) The band structure with $t'=0.4t$. (b) The same band structure along the $k_x$-axis.}\label{Fig2}
\end{figure}

\begin{figure}[t]
  \centering
  \subfigure[]{\includegraphics[width=4cm]{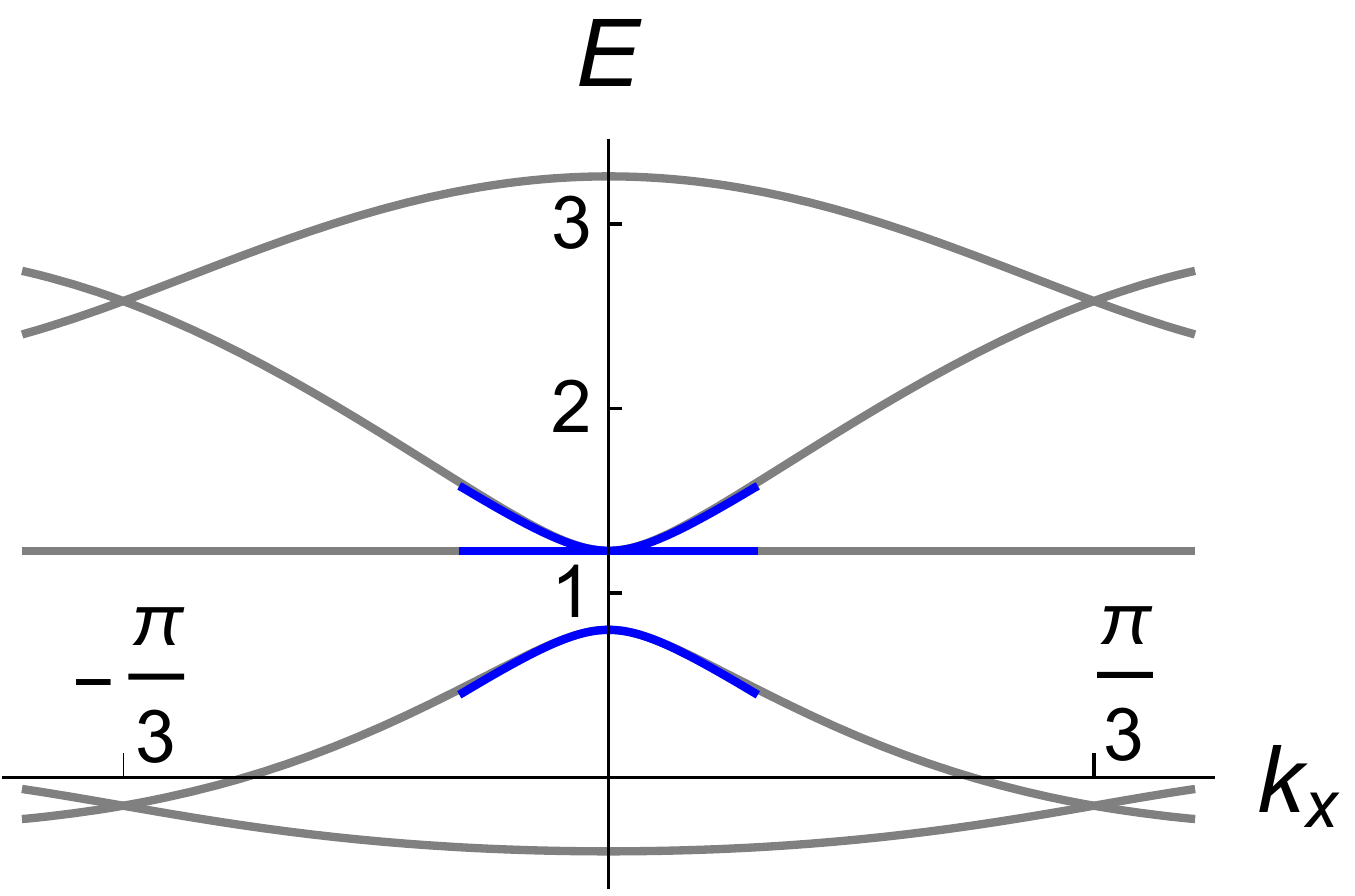}\label{5band1}}~~
  \subfigure[]{\includegraphics[width=4.cm]{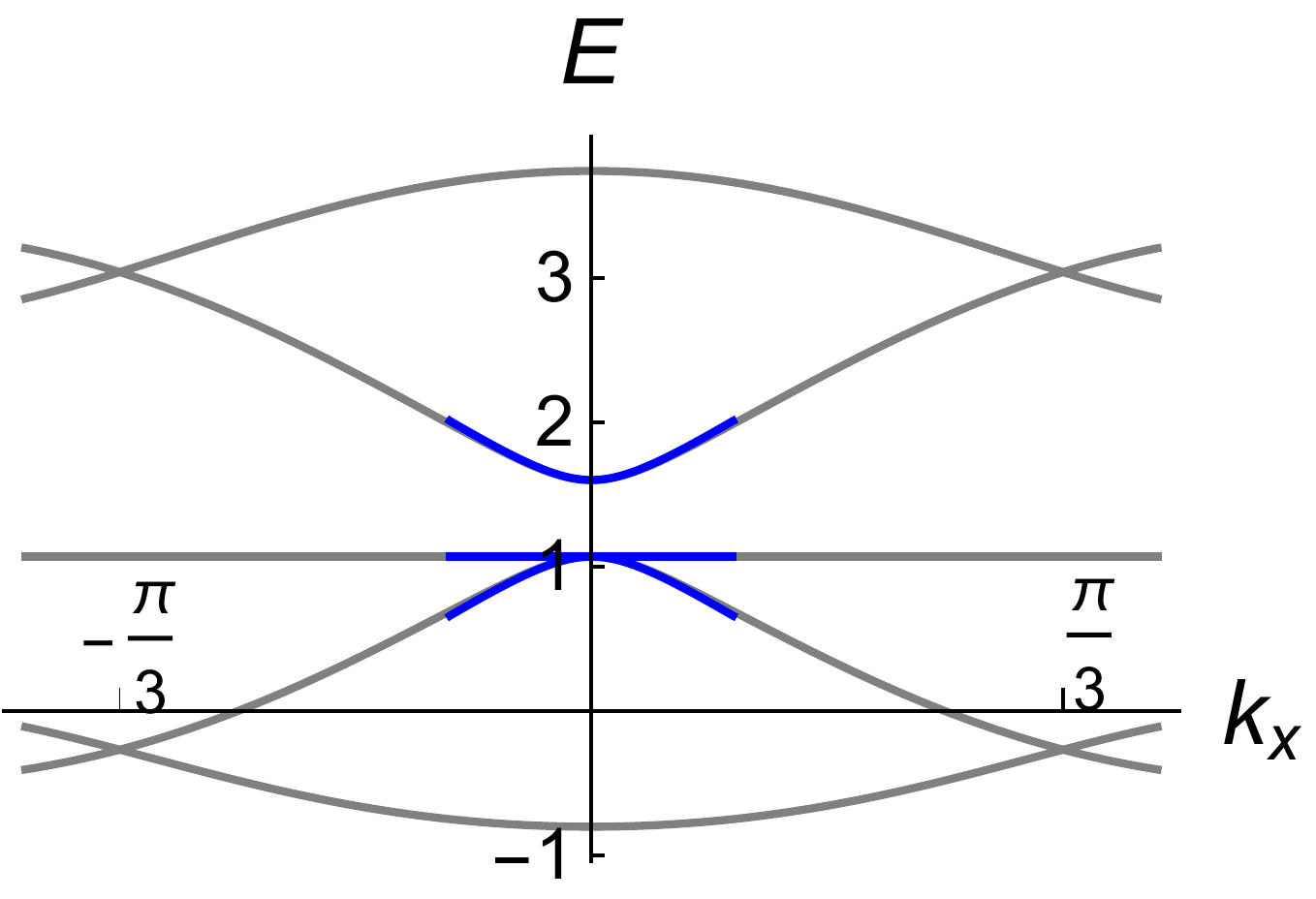}\label{5band3}}
  \subfigure[]{\includegraphics[width=4cm]{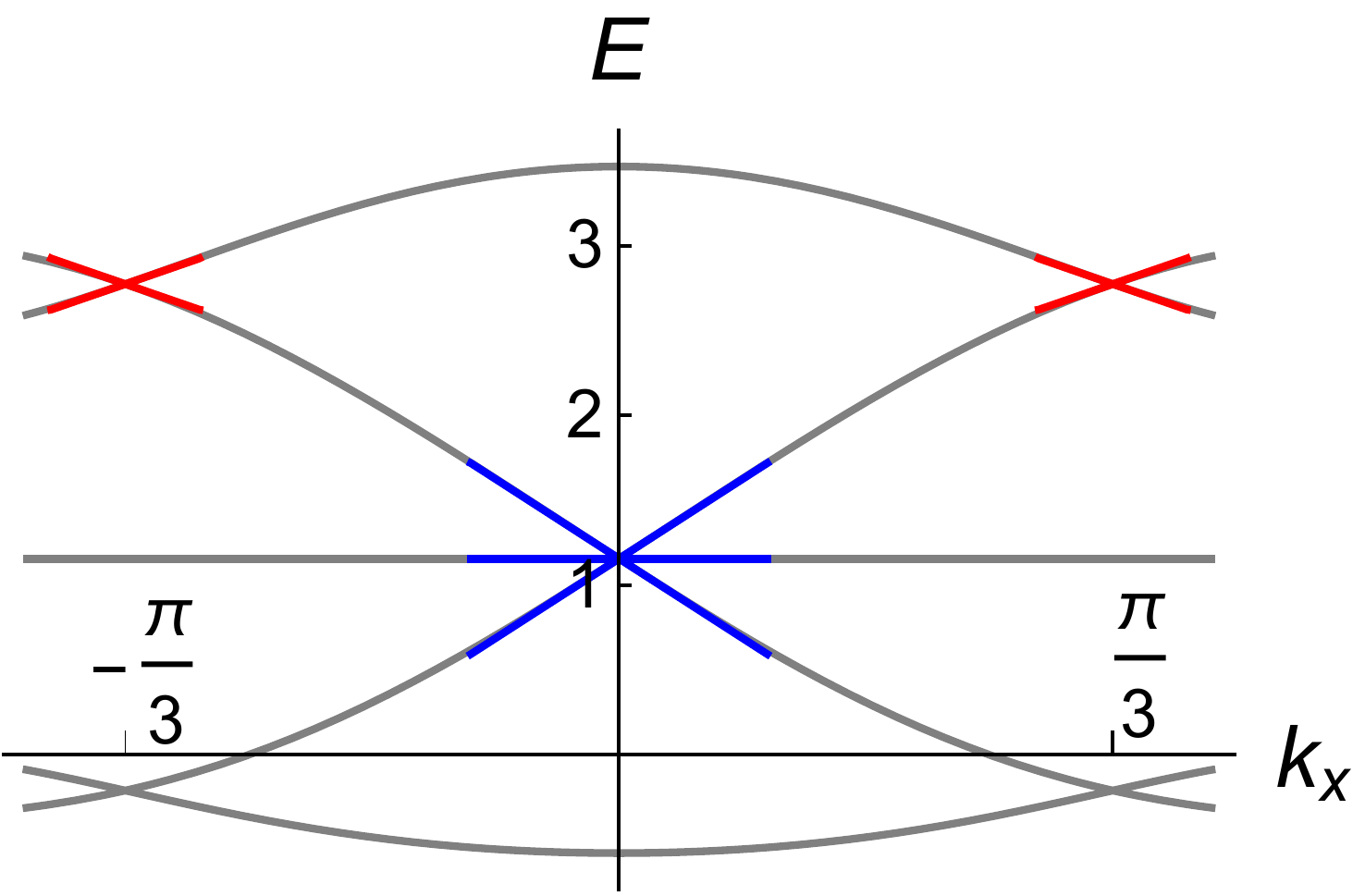}\label{5band2}}
  \subfigure[]{\includegraphics[width=4.5cm]{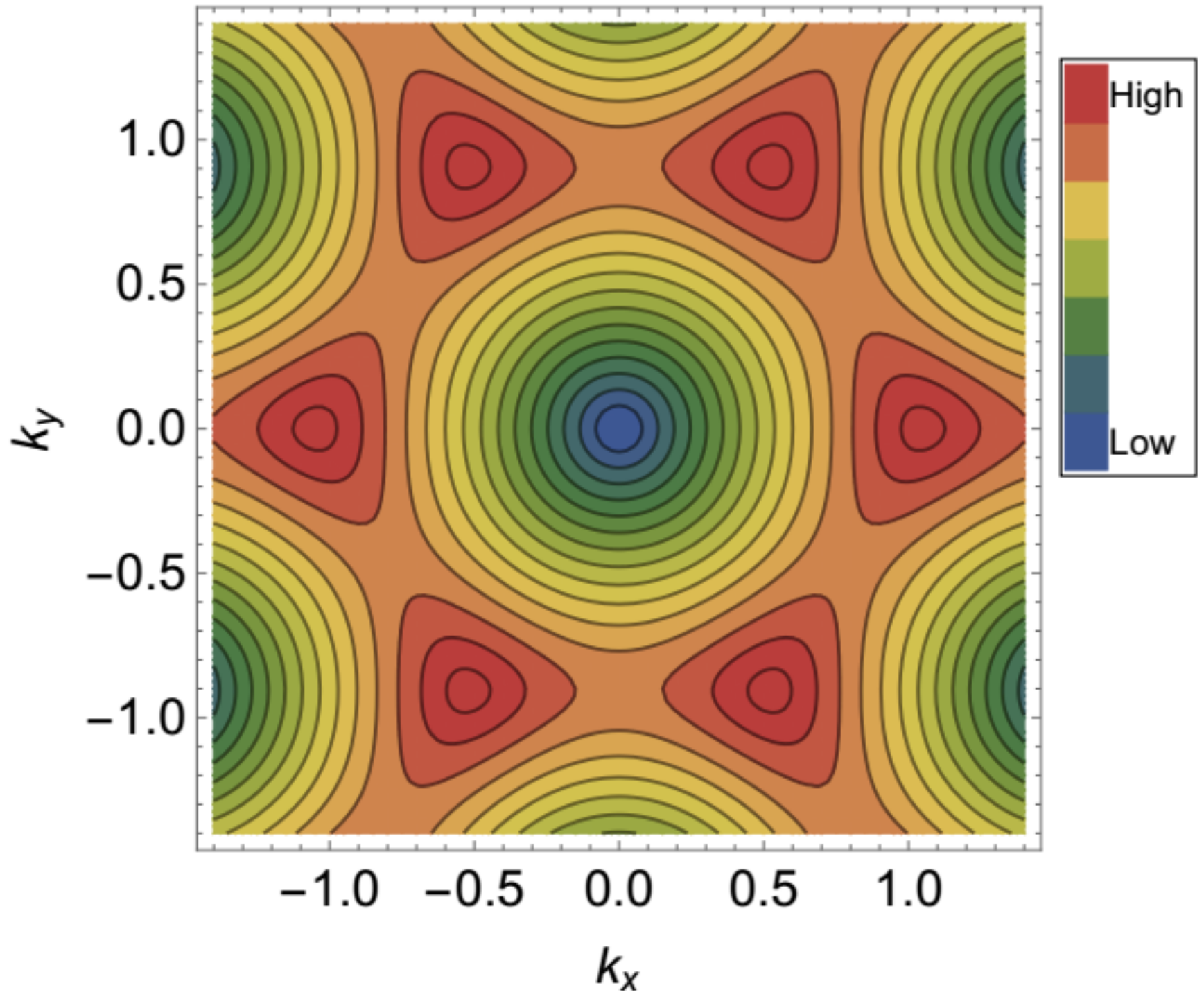}\label{ConstE}}
  \caption{(a)-(c) The upper five bands along the $k_x$-axis with (a) $t'=0.4t$, (b) $t'=0.8t$ and (c) $t'\approx 0.579t$. The spectrum in blue is that of the spin-1 Hamiltonian with a ``mass" term in (a) and (b) and without ``mass" terms in (c), and the spectrum in red in (c) is that of the Dirac Hamiltonian. (d) The constant energy contours of the second highest band at $t'\approx 0.579t$. }\label{Fig3}
\end{figure}

{\it Dirac fermion and spin-1 fermion on the TKL}.---A schematic of the TKL is shown in Fig.\ref{Fig1}. It can be viewed as a kagome lattice decorated with an additional triangle inside each of its original triangles. Let the distance between the nearest neighbors be $a$. We choose the primitive vectors $\vec{a}_1=(4a,0)$ and $\vec{a}_2=(-2a,2\sqrt{3}a)$, and a unit cell is indicated by the red parallelogram in Fig.\ref{Lattice}. We will set $a=1$ for simplicity. The TKL has $D_{6h}$ point group symmetry, as can be seen from the part enclosed by the blue rectangle in Fig.\ref{Lattice}. The first Brillouin zone is shown in Fig.\ref{BZ}, with three high symmetry momenta labeled: ${\bf\Gamma}=(0,0)$, and the two inequivalent Brillouin zone corners $\bK=(\frac{\pi}{3},0)$ and $\bK'=(-\frac{\pi}{3},0)$. We study the tight-binding Hamiltonian $\mathcal{H}=\sum_{\langle ij\rangle}t_{ij}c_i^\dagger c_j$ and only nearest neighbor hopping is considered. While the kagome lattice has three sites in each unit cell, the TKL has nine, resulting in nine bands if only a single orbit on each site is considered. We assume the hopping energy is $t$ and $t'$ for the solid bonds and dashed bonds in Fig.\ref{Lattice}, respectively. A typical band structure is shown in Fig.\ref{Fig2}, with $t'=0.4t$. Generically, the nine bands decouple into three groups, which we call the upper, middle and lower group according to their energy range. There are two Dirac points at ${\bf K}$ and ${\bf K}'$ in each group of bands. We find that three of the nine bands are flat, each touching with another band. The flatness of those bands is due to the frustrated hopping which yields fully localized Wannier states, and the band touching can be understood from a state counting argument\cite{Bergman2008}. Upon tuning the ratio $t'/t$, the flat band can be shifted between different groups. We will focus on the upper and middle group from here on, and we show only the five upper bands in Fig.\ref{5band1} to \ref{5band2} for different parameters. As can be seen, at $t'<t_c$ where $t_c\approx0.579t$, each group has one flat band, while at $t'>t_c$, the flat band of the upper group is shifted to the middle group. During the process of tuning $t'/t$, there is a critical point $t'=t_c$ where three bands touch at ${\bf\Gamma}$ point, as shown in Fig.\ref{5band2}. We demonstrate below that near this band touching, the 2D fermions are governed by the spin-1 Hamiltonian. The constant energy contours of the second highest band are plotted in Fig.\ref{ConstE}. At low energy, circular electron pockets around ${\bf\Gamma}$ point originate from the spin-1 Hamiltonian. Hexagonal warping appears at higher energy due to the $D_{6h}$ symmetry of ${\bf\Gamma}$ point. At high energy, circular hole pockets around $\bK$ and $\bK'$ originate from the Dirac Hamiltonian, and trigonal warping appears due to the $C_{3v}$ symmetry of $\bK$ and $\bK'$, the same as the case in graphene\cite{Neto2009}.

We use $k\cdot p$ method to find the effective Hamiltonian near the threefold degeneracy. The tight-binding Hamiltonian can be written as $\mathcal{H}=\sum_{\bk\in B.Z.}\Psi^\dagger(\bk) H(\bk)\Psi(\bk)$ where $\Psi(\bk)$ is a nine-component spinor. First we diagonalize $H(\bk)$ at ${\bf\Gamma}$ point, and find the eigenenergies $E_i(0)$ in ascending order from the lowest to the highest, with the associated wave functions $|\psi_i(0)\rangle$. Then we expand $H(\bk)$ around ${\bf\Gamma}$ point to the first order of $\bk$, and the effective Hamiltonian near the triple degeneracy $H_3(\bk)$ has elements $[H_3(\bk)]_{ij}=\langle\psi_{i-5}(0)|H(\bk)|\psi_{j-5}(0)\rangle$ with $i,j=6,7,8$. An arbitrarily chosen set of orthogonal basis results in $H_3(\bk)=\eps+\sum_{i=x,y;j=x,y,z}k_iv_{ij}S_j$, where we choose three out of the eight Gell-Mann matrices as the spin-1 matrices
\begin{eqnarray}
  S_x&=&\left(\begin{array}{ccc}0&0&i\\0&0&0\\
   -i&0&0\end{array}\right),
  S_y = \left(\begin{array}{ccc}0&0&0\\0&0&i\\
   0&-i&0\end{array}\right),\nonumber\\
  S_z&=&\left(\begin{array}{ccc}0&-i&0\\i&0&0\\
   0&0&0\end{array}\right)
\end{eqnarray}
which satisfy the angular momentum algebra $[S_i,S_j]=i\eps_{ijk}S_k$ where $\eps_{ijk}$ is the Levi-Civita symbol. We can choose a specific set of basis so that the velocity matrix has elements $v_{ij}=v\delta_{ij}$, and then the effective Hamiltonian is simply
\begin{eqnarray}
   H_3(\bk)=\eps+v\bk\cdot{\bf S}=\eps+v(k_xS_x+k_yS_y).
\end{eqnarray}
This Hamiltonian describes a 2D spin-1 fermion, with the eigenenergies $\eps,\eps\pm v\sqrt{k_x^2+k_y^2}$. The spin-1 spectrum is shown in blue in Fig.\ref{5band2}, and can be viewed as a Dirac cone touching a flat band. Different from spin-1/2 fermions, a spin-1 fermion can exist on its own and avoids the fermion-doubling theorem by Nielsen and Ninomiya\cite{Nielsen1981}.

Near the double degeneracy at $\bK$ point, we use the same method to find the effective Hamiltonian $H_{\bK}(\delta\bk)=\eps_1+v_D\delta\bk\cdot{\bf \sigma}$, where $\delta\bk$ is the momentum deviation from $\bK$, and ${\bf\sigma}_i$'s are Pauli matrices. So this is a Dirac point in 2D. The Hamiltonian near the other Dirac point $\bK'$ is related to $H_{\bK}$ by the time-reversal operation $\Theta=K$ where $K$ is the complex conjugation, $H_{\bK'}(\delta\bk)=\eps_1-v_D\delta\bk\cdot{\bf\sigma^*}$. The spectrum of the Dirac Hamiltonian is shown in red in Fig.\ref{5band2}.

We have therefore shown that spin-1 fermion and spin-1/2 Dirac fermion could coexist on the TKL if the ratio of $t'/t$ is fine tuned to a critical value ($t_c\approx0.579t$). Away from the critical value, a ``mass" term of the form
\begin{eqnarray}
\left(\begin{array}{ccc}\Delta&0&0\\0&0&0\\
   0&0&0\end{array}\right)
\end{eqnarray}
appears and opens a gap between one dispersive band and the band touching. Depending on the sign of $\Delta$, it could be either the case as in Fig.\ref{5band1} or that as in Fig.\ref{5band3}.

{\it Effective model with both Dirac and spin-1 fermions}.---An effective model which hosts different types of fermions is highly desired to study their interplay. Here, we derive a four-band effective model with both Dirac fermions and spin-1 fermion from the tight-binding Hamiltonian of the TKL. Note that an effective model including the uppermost four bands could have both types of fermions. Using the $k\cdot p$ method as in the above section, we find the Hamiltonian of the uppermost four bands near ${\bf\Gamma}$ point $H_4(\bk)=$
\begin{eqnarray}
  \left(\begin{array}{cccc}0&0&ivk_x&-2\al k_xk_y\\0&0&ivk_y&\al(k_y^2-k_x^2)\\
   -ivk_x&-ivk_y&0&0\\ -2\al k_xk_y&\al(k_y^2-k_x^2)&0&\eps'-\frac{k^2}{2m}\end{array}\right),
\end{eqnarray}
where $k=\sqrt{k_x^2+k_y^2}$. We have kept only linear terms in the upper left $3\times3$ block, but up to quadratic terms otherwise. The reason is that we would like to derive a minimal model for the coexistence of the two types of fermions, whereas including quadratic terms in the upper left block would complicate the model. If $\al=0$, the band structure of $H_4$ is simply a spin-1 cone intersecting with a quadratic band dispersing downward (parameterized by $\eps'>0$ and $m>0$) along a ring. The terms linear in $\al$ gap the degeneracy of the nodal ring, leaving only discrete Dirac points, as we explain later.

The four-band effective Hamiltonian can be written in a more compact form,
\begin{eqnarray}
  H_4(\bk) &=& \frac{1}{2}v(k_+S_-+k_-S_+)+(\eps'-\frac{k_+k_-}{2m})M\nonumber\\
  &+&\frac{1}{2}\al(k_+^2L_++k_-^2L_-),\label{eq:H4}
\end{eqnarray}
where we have defined $k_\pm=k_x\pm ik_y$, $S_\pm=S_x\pm iS_y$ and $L_\pm=L_x\pm iL_y$, in which
\begin{eqnarray}
  S_x&=&\left(\begin{array}{cccc}0&0&i&0\\0&0&0&0\\
   -i&0&0&0\\0&0&0&0\end{array}\right),
  S_y = \left(\begin{array}{cccc}0&0&0&0\\0&0&i&0\\
   0&-i&0&0\\0&0&0&0\end{array}\right),\nonumber\\
  S_z&=&\left(\begin{array}{cccc}0&-i&0&0\\i&0&0&0\\
   0&0&0&0\\0&0&0&0\end{array}\right),
  L_x = \left(\begin{array}{cccc}0&0&0&0\\0&0&0&-1\\
   0&0&0&0\\0&-1&0&0\end{array}\right),\nonumber\\
  L_y&=&\left(\begin{array}{cccc}0&0&0&1\\0&0&0&0\\
   0&0&0&0\\1&0&0&0\end{array}\right),
  L_z = \left(\begin{array}{cccc}0&0&0&0\\0&0&0&0\\
   0&0&0&-1\\0&0&-1&0\end{array}\right),\nonumber\\
  M&=&\left(\begin{array}{cccc}0&0&0&0\\0&0&0&0\\
   0&0&0&0\\0&0&0&1\end{array}\right).
\end{eqnarray}
The $S_i$ and $L_i$ are generators of SO(4) group which satisfy $[S_i,S_j]=i\eps_{ijk}S_k$, $[L_i,L_j]=i\eps_{ijk}S_k$ and $[S_i,L_j]=i\eps_{ijk}L_k$. The rotational operation is represented as $e^{iS_z\theta}$. Under $C_3$ operation and time-reversal operation, the momentum and the matrices are transformed as
\begin{eqnarray}
  C_3&:&k_\pm\rightarrow e^{\pm i2\pi/3}k_\pm, S_\pm\rightarrow e^{\pm i2\pi/3}S_\pm, L_\pm\rightarrow e^{\pm i2\pi/3}L_\pm,\nonumber\\
  &&S_z\rightarrow S_z, L_z\rightarrow L_z, M\rightarrow M\nonumber\\
  \Theta&:&\bk\rightarrow -\bk, \bS\rightarrow -\bS, \bL\rightarrow \bL, M\rightarrow M.\label{eq:trans}
\end{eqnarray}
Then it is clear that $H_4(\bk)$ is invariant under Eq.\ref{eq:trans}. Although $H_4(\bk)$ has threefold rotational symmetry, the band structure is sixfold rotational invariant due to the time-reversal symmetry.

\begin{figure}[t]
  \centering
  \subfigure[]{\includegraphics[width=4.cm]{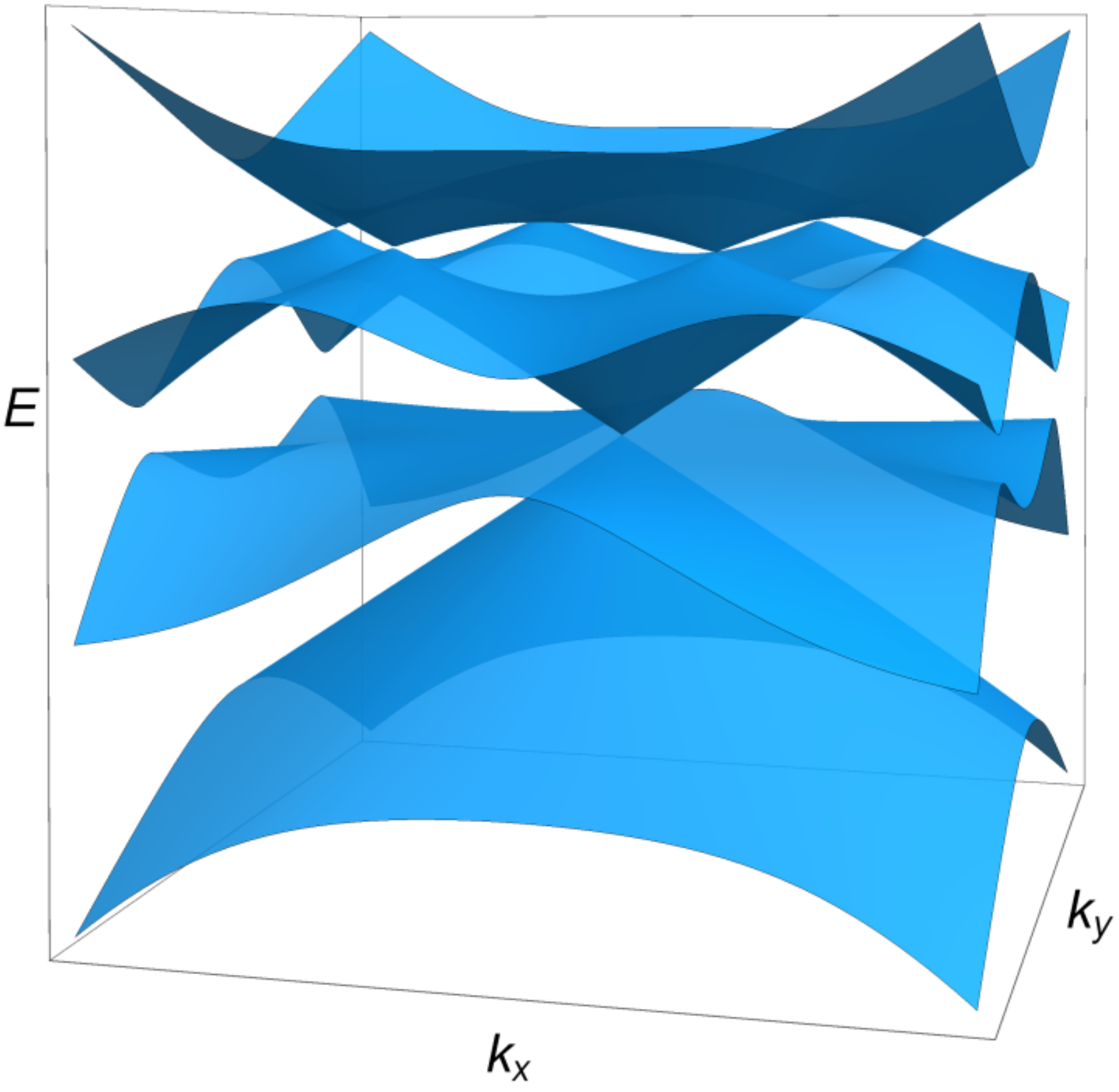}\label{4band}}~~
  \subfigure[]{\includegraphics[width=4.5cm]{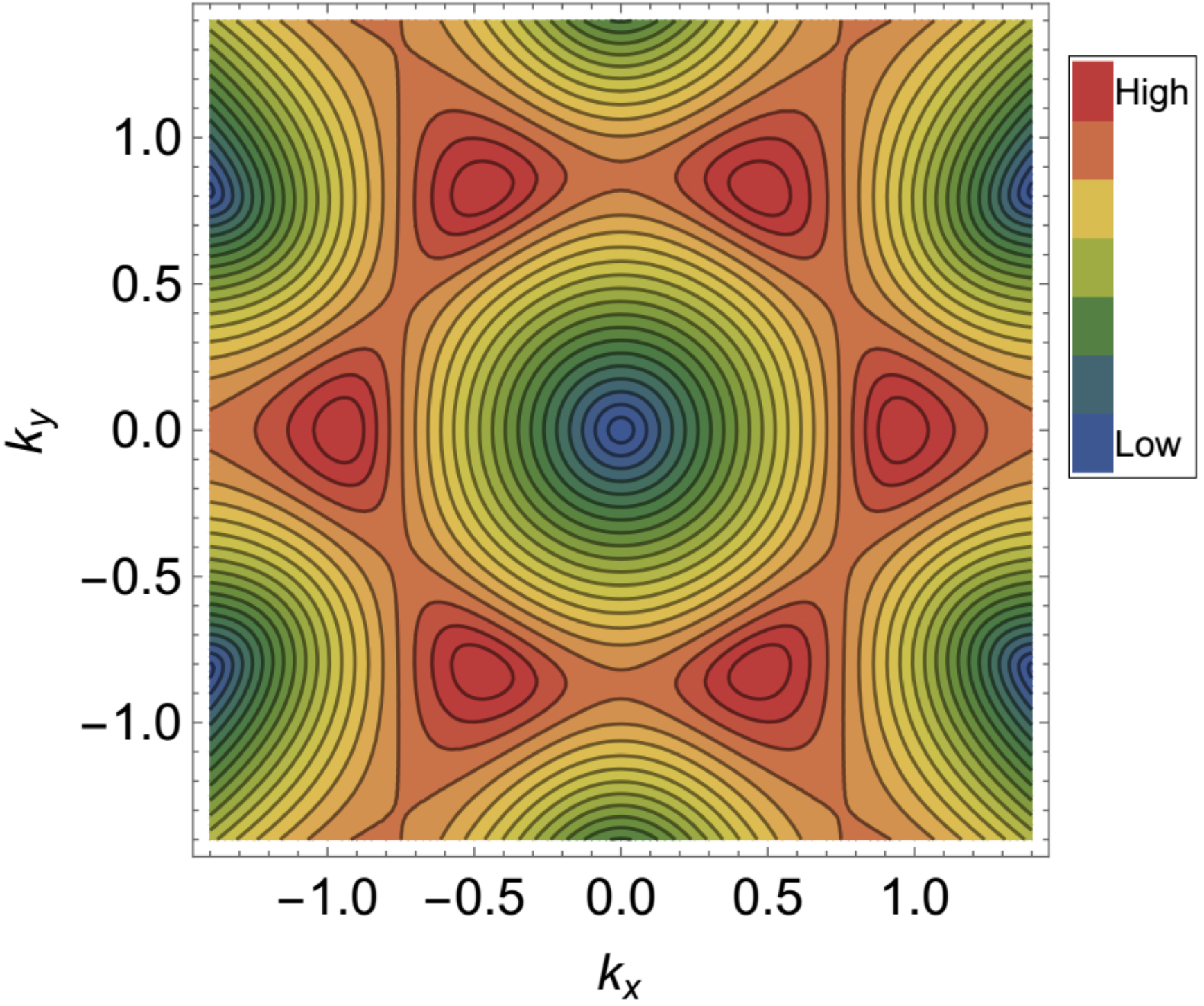}\label{ConstEn2}}
  \caption{(a) Band structure of the four-band model Eq.\ref{eq:H4}. (b) The constant energy contours of the second highest band of the four-band model Eq.\ref{eq:H4}. The parameters are $v=1.78$, $\al=0.5$, $\eps'=2.31$ and $m=0.58$.}\label{Fig4}
\end{figure}

We show the band structure of $H_4(\bk)$ in Fig.\ref{4band}, and constant energy contours in Fig.\ref{ConstEn2}. The spin-1 cone and Dirac cones are well reproduced, and hexagonal warping to the spin-1 cone also appears because of the symmetries mentioned above. However, due to the absence of rotational symmetry at each Dirac point, the Dirac cones are in general anisotropic and tilted, and the warping has not threefold rotational symmetry anymore.

The full expression of the spectrum is complicated, and is devoid of a simple interpretation of the appearance of the Dirac points. A better understanding of the coexistence of spin-1 and Dirac fermions can be achieved if we treat the term linear in $\al$ in Eq.\ref{eq:H4} as a perturbation, and write the Hamiltonian in two parts: $H_4=H_4^{(0)}+H_4^{(1)}$ where $H_4^{(0)}=\frac{1}{2}v k(e^{i\theta}S_-+e^{-i\theta}S_+)+(\eps'-\frac{k^2}{2m})M$ and $H_4^{(1)}=\frac{1}{2}\al k^2(e^{2i\theta}L_++e^{-2i\theta}L_-)$, with $\theta\equiv\tan^{-1}k_y/k_x$. The four bands of $H_4^{(0)}$ have eigenenergies $E_1^{(0)}=\eps'-\frac{k^2}{2m}$, $E_2^{(0)}=vk$, $E_3^{(0)}=0$ and $E_4^{(0)}=-vk$, with the associated wave functions $|1^{(0)}\rangle=\{0,0,0,1\}^t$, $|2^{(0)}\rangle=\frac{1}{\sqrt{2}}\{i\cos\theta,i\sin\theta,1,0\}^t$, $|3^{(0)}\rangle=\{-\sin\theta,\cos\theta,0,0\}^t$ and $|4^{(0)}\rangle=\frac{1}{\sqrt{2}}\{-i\cos\theta,-i\sin\theta,1,0\}^t$, respectively. The first two bands touch at a nodal ring given by $k^*=mv(\sqrt{1+\frac{2\eps'}{mv^2}}-1)$. Transforming $H_4^{(1)}$ to the eigenbasis of $H_4^{(0)}$, then the effective Hamiltonian for the upper two bands is
\begin{eqnarray}
  H_2(\bk)&=&\frac{1}{2}(\eps'-\frac{k^2}{2m}+vk)+\frac{1}{2}(\eps'-\frac{k^2}{2m}-vk)\si_z\nonumber\\
  &+&\frac{1}{\sqrt{2}}\al k^2\sin 3\theta\si_y.
\end{eqnarray}
Obviously, the last term opens a gap along the nodal ring except at six points, $\theta^*=n\pi/3$ with $n=0,1,...5$. Since the spectrum has sixfold rotational symmetry, we expand $H_2(\bk)$ near one of the points, $\theta^*=0$, i.e. $\bk=(k^*,0)$, and get
\begin{eqnarray}
  H_2^{eff}(\delta k_x,k_y)&=&vk^*+v_x'\delta k_x+v_x\delta k_x\si_z+v_y k_y\si_y,
\end{eqnarray}
where $v_x'=\frac{1}{2}(v-\frac{k^*}{m})$, $v_x=-\frac{1}{2}(v+\frac{k^*}{m})$, and $v_y=\frac{3}{\sqrt{2}}\al k^*$. $H_2^{eff}$ is, for generic parameters, the Hamiltonian of a tilted anisotropic Dirac cone. The condition for the Dirac cone to be untilted and isotropic is $v_x'=0$ and $v_x=v_y$, which yields $\eps'=\frac{3}{2}mv^2$ and $m\al=\frac{\sqrt{2}}{3}$. Nevertheless, the parameters derived from the tight-binding Hamiltonian $H(\bk)$ do not have to satisfy this condition, since the four-band Hamiltonian $H_4$ is a $k\cdot p$ Hamiltonian around ${\bf \Gamma}$ point, hence, in general, it does not capture the exact band structure around $\bK$ point.

\begin{figure}[t]
  \centering
  \subfigure[]{\includegraphics[width=5.cm]{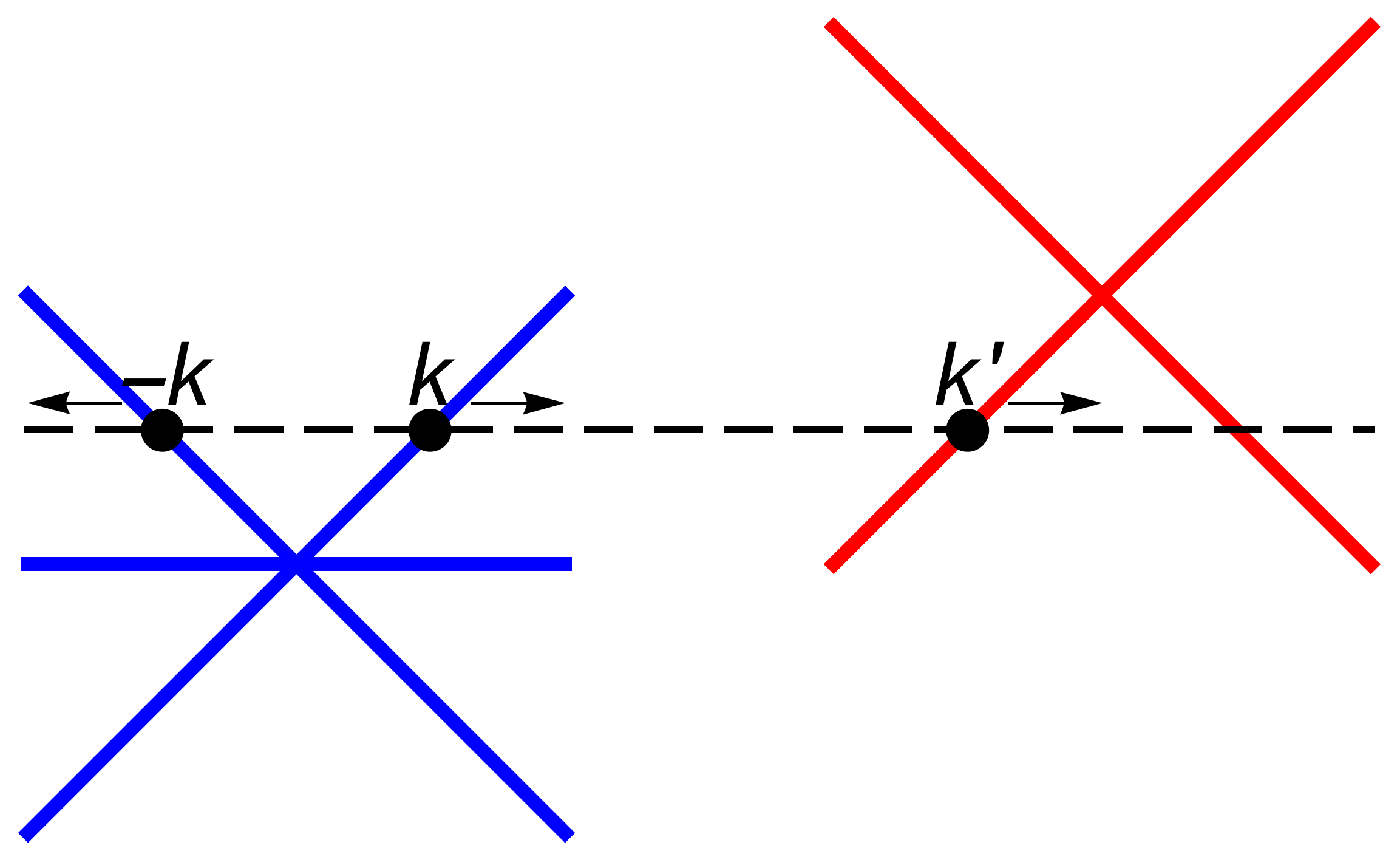}\label{KTdemon}}
  \subfigure[]{\includegraphics[width=8cm]{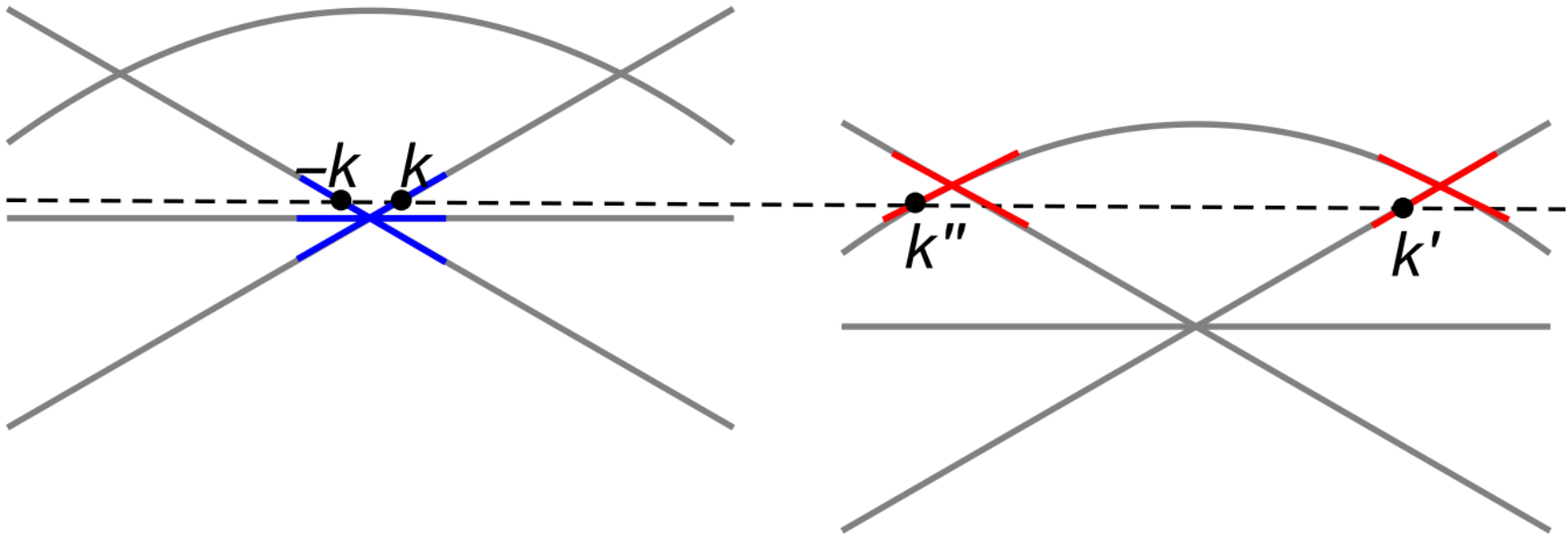}\label{KT}}
  \caption{(a) Schematic of spin-nonconserving Klein tunneling from spin-1 to spin-1/2 fermion. (b) Spin-nonconserving Klein tunneling when a potential step with a proper depth is present.}\label{Fig5}
\end{figure}

{\it Spin-nonconserving Klein tunneling}.---The conventional Klein tunneling occurs when a relativistic particle is incident on a high potential barrier. It has been studied in the context of graphene\cite{Katsnelson2006}, and verified in several experiments\cite{Stander2009,Young2009,Young2011}. The Klein tunneling of spin-1 fermions has also been addressed\cite{Urban2011,Xu2014}. The effective model Eq.\ref{eq:H4} is a good start to study the interplay between spin-1 and spin-1/2 fermions in lattice systems. As shown in Fig.\ref{Fig5}, a fermion can tunnel from a spin-1 particle state with momentum ${\it \bk}$ to a spin-1/2 hole state with momentum $\bk'$ or $\bk''$ when there is a potential step with a proper depth, which we name as the spin-nonconserving Klein tunneling. The perturbed wave functions of the upper two bands are, respectively, $|1\rangle=|1^{(0)}\rangle+\frac{i\al k^2\sin 3\theta}{\sqrt{2}(\eps'-k^2/2m-vk)}|2^{(0)}\rangle- \frac{\al k^2\cos3\theta}{\eps'-k^2/2m}|3^{(0)}\rangle-\frac{i\al k^2\sin 3\theta}{\sqrt{2}(\eps'-k^2/2m+vk)}|4^{(0)}\rangle$ and $|2\rangle=|2^{(0)}\rangle+\frac{i\al k^2\sin 3\theta}{\sqrt{2}(\eps'-k^2/2m-vk)}|1^{(0)}\rangle$. At normal incidence, $\theta=0$, the wave functions on the left side and right side of the potential step are
\begin{eqnarray}
  \psi_L(x) &=& Ae^{ikx}\left(\begin{array}{c}
                                  i \\
                                  0 \\
                                  1 \\
                                  0
                                \end{array}\right)+Be^{-ikx}\left(\begin{array}{c}
                                  -i \\
                                  0 \\
                                  1 \\
                                  0
                                \end{array}\right),\\
  \psi_R(x) &=& Ce^{ik'x}\left(\begin{array}{c}
                                  i \\
                                  0 \\
                                  1 \\
                                  0
                                \end{array}\right)+De^{ik''x}\left(\begin{array}{c}
                                  0 \\
                                  0 \\
                                  0 \\
                                  1
                                \end{array}\right),
\end{eqnarray}
respectively. From the continuity of the wave function at $x=0$, we find $A=C$ and $B=D=0$. Therefore, through the spin-nonconserving Klein tunneling, a spin-1 particle tunnels to a spin-1/2 hole state near $\bK$ with the transmission coefficient $T=1$. Veselago lens and transistors based on Klein tunneling in graphene have been conceived\cite{cheianov2007,Wilmart2014}, and similar devices based on spin-nonconserving Klein tunneling are expected.

{\it Discussion}.---The (breathing) kagome lattice which has three sites in each unit cell can host spin-1 fermion or Dirac fermions, but not their coexistence\cite{Essafi2017}, since it has only three bands and one of them is flat. Compared with it, the TKL has the merit of having more bands while retaining the same lattice symmetry, hence can host both types of fermions. Nevertheless, as we have shown, a four-band model is capable of doing this, thus a lattice with fewer sites in each unit cell and with $D_{6h}$ symmetry may achieve this goal. This is left for future study.

Since the band structure of a lattice does not depend on the statistics of the quasiparticles under study, there also exist spin-1/2 Weyl photons and Weyl magnons as well as spin-1 photons in periodic systems. Therefore, the TKL can also act as a platform for the coexistence of spin-1 bosons and spin-1/2 bosons in 2D. Experimentally, ultracold atoms in a tunable optical TKL could be designed to study the interplay between the two types of fermions or bosons.

More interesting questions based on the tight-binding model of the TKL and the effective model are awaiting further studies, such as the magnetotransport properties and the topological invariant of each band once the fermions are gapped by spin-orbit coupling. The extension to three dimensions will also be an exciting direction.

{\it Acknowledgement}. L.W. and D.X.Y. are supported by NKRDPC-2017YFA0206203, NSFC-11574404, NSFG-2015A030313176, National Supercomputer Center in Guangzhou, and Leading Talent Program of Guangdong Special Projects.

\bibliography{TKLbib}
\end{document}